# Characterizing Halloumi cheese's bacterial communities through metagenomic analysis

**Authors:** Eleni Kamilari, MSc[1], Dimitrios A. Anagnostopoulos[1], Photis Papademas[1], Andreas Kamilaris[2, 3], Dimitris Tsaltas[1].

**Affiliations:**

1. Department of Agricultural Sciences, Biotechnology and Food Science, Cyprus University of Technology, Limassol, Cyprus

2. Research Centre on Interactive Media, Smart Systems and Emerging Technologies (RISE), Nicosia, Cyprus

3, Department of Computer Science, University of Twente, the Netherlands

**Abstract**

Halloumi is a semi-hard cheese produced in Cyprus for centuries and its popularity has significantly risen over the past years. High-throughput sequencing (HTS) was applied in the present research to characterize traditional Cyprus Halloumi bacterial diversity. Eighteen samples made by different milk mixtures and produced in different areas of the country were analyzed, to reveal that Halloumi's microbiome was mainly comprised by lactic acid bacteria (LAB), including *Lactobacillus, Leuconostoc*, and *Pediococcus,* as well as halophilic bacteria, such as *Marinilactibacillus* and *Halomonas.* Additionally, spore forming bacteria and spoilage bacteria, were also detected. Halloumi produced with the "traditional" method, had significantly richer bacterial diversity compared to Halloumi produced with the "industrial" method. Variations detected among the bacterial communities highlight the contribution of the initial microbiome that existed in milk and survived pasteurization, as well as factors associated with Halloumi manufacturing conditions, in the final microbiota composition shaping. Identification and characterization of Halloumi microbiome provides an additional, useful tool to characterize its typicity and probably safeguard it from fraud products that may appear in the market. Also, it may assist producers to further improve its quality and guarantee consumers' safety.

## 1. Introduction

Halloumi cheese originated from Cyprus and has been produced on the island for centuries (Welz G., 2015). Nowadays due to its great popularity" Halloumi-type" cheese is produced in low-quantities in the region known as "Levant", including Lebanon, Syria, and Turkey, but also in other countries, such as United Kingdom, Bulgaria, and Sweden. Halloumi cheese production follows the standard CYS 94 (Parts 1,2 for fresh and mature Halloumi cheese, respectively) which was set in 1985 by Government bodies (CYS 94, 1985). As described by Papademas and Robinson (Papademas & Robinson, 1998), traditional fresh Halloumi is made by fresh pasteurized milk, which is coagulated by rennet, and after the "cooking" step , the cheese is left to briefly cool down, manually folded, salted, sprinkled with dry *Mentha viridis* leaves and stored overnight in 11-12% NaCl whey until the individual halloumi cheese pieces (250gr) are vacuum-packed and kept at 6°C until consumption. For mature Halloumi cheese production, the fresh cheese is stored in the whey brine (11-12%) NaCl for 40 days at 15–20°C, before vacuum packed. Regarding the type of milk, the standard CYS 94 (CYS 94, 1985) indicates that goat and/or sheep milk should be used but accepts the presence of cow milk.

Halloumi is characterized by unique organoleptic characteristics arising from the specific methods of production, as well as by the contribution of the indigenous microflora that exists in the milk (Papademas & Robinson, 1998). The application of thermal process and the absence of starter culture, indicates that the existing microbiome is affected by several factors, including: a) milk pasteurization and cooking procedures, b) microbial contamination by rennet, as well as salt and *Mentha viridis* leaves addition, c) microbial contamination from dairy unit environment and d) the preservation in brined whey (CYS 94, 1985; Calasso et al., 2016; De Pasquale et al., 2016; Delcenserie et al., 2014; Stellato et al., 2015; Yeluri Jonnala et al., 2018). Culture-based techniques identified the presence of spore forming bacteria, such as *Bacillus*, as well as thermophilic species, such as members of the genus *Lactobacillus* and *Enterococcus*, in addition to yeasts, in Halloumi samples (Bintsis & Papademas, 2002; Tamime, 2007). In 2001, phenotypic and phylogenetic analyses led to the characterization of a novel, salt-tolerant species in Halloumi samples, named *Lactobacillus cypricasei* sp. nov., (Lawson et al., 2001), which was later found to be heterotypic synonym of *Lactobacillus acidipiscis* (Kim et al., 2011; Naser, Vancanneyt, Hoste, Snauwaert, & Swings, 2006; Sun et al., 2015).

Despite Halloumi's high consumption globally, the autochthonous microbial communities that affect its unique organoleptic properties and are associated with the product's safety for consumers, have not yet been evaluated. The application of High Throughput Sequencing (HTS) technology will facilitate the identification of the complex microbial communities of Cyprus Halloumi. Amplicon sequencing technology enables comprehensive characterization of the microbiota within a sample with higher sensitivity and higher throughput detection, compared to other molecular techniques through massive parallel sequencing of small fragments of universally conserved DNA sequences, such as the bacterial 16S rRNA gene (Bokulich & Mills, 2013). Due to the capabilities of microbial communities to generate thousands of reading sequences, this fact may provide the potential to characterize them, including even low-abundance bacteria. Additional advantages of applying the current technology for the identification and characterization of the microbial communities in Protected Denomination of Origin (PDO) cheeses has been reviewed recently (Kamilari et al. 2019). The microbiome of Halloumi can be expected to be diverse among producers due to variations in the microbial composition of the milk, differences associated with the manufacturing conditions, including geographical area, microbiota of the dairy plant environment and animal breed of the milk (Bokulich and Mills, 2013; Dalmasso et al., 2016; Yeluri Jonnala et al., 2018; Kamilari et al. 2019). The main contribution of this paper is to reveal the bacterial communities of Cyprus Halloumi, as sold in the Cypriot market, produced by small or large industries, with or without the addition of cow milk, following the CYS 94 (1985) standard. Microbiome analysis has been achieved using Illumina MiSeq amplicon sequencing. The extracted findings may enable the characterization of the microbiome developed in Cyprus Halloumi and assist to the improvement of the product quality and safety for consumers.

Importantly, the current analysis provides indications whether the microbiome may be used as an additional tool in order to define the typicity of Cyprus Halloumi.

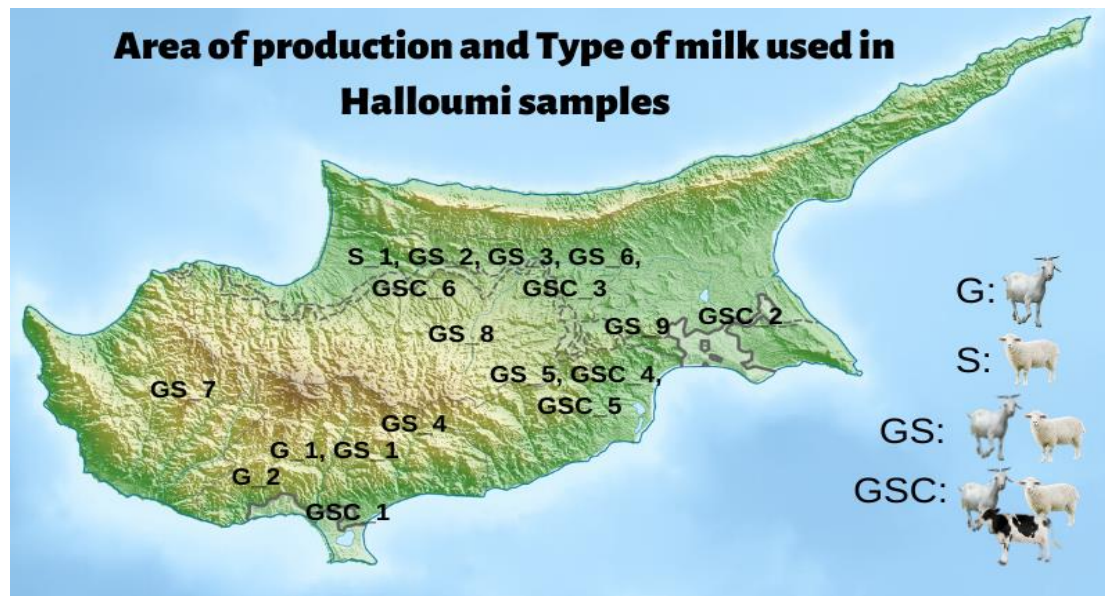

**Figure 1**. Map of Cyprus showing areas of production and type of milk used in Halloumi samples.

## 2. Methods

### 2.1 Sample collection

Halloumi samples were collected from production sites (dairies) and the market. Specifically, five samples were collected from the production sites and thirteen were collected from supermarkets, with the criteria: a) to be produced in different geographic areas of the Republic of Cyprus and b) being composed by goat and/or sheep milk (considered as "traditional"), or a mixture of goat, sheep and cow milk (considered as "industrial") (see Figure 1, Table 1). The two manufacturing methods are furthermore differentiated by harsher heat treatments performed by large industries in order to extend the shelf life of the product kept after refrigerated conditions. From these samples twelve were made only by goat or sheep milk or by mixture of goat and sheep milk and six by a mixture of goat, sheep and cow milk. Sixteen of the Halloumi samples were fresh and two were matured (40 days ripened). Samples were transported in cool conditions (ice packs) and stored at -20₀ C until processing.

### 2.2 Metagenomic DNA extraction

For sample homogenization, twenty grams of Halloumi were homogenized in 180 ml of 2% tri-sodium citrate (Honeywell, Europe) using Stomacher 400 Circulator (Seward, UK) at 300 rpm for 2 minutes. DNA extraction was performed using DNeasy® PowerFood® Microbial Kit (MoBio Laboratories Inc., Carlsbad, CA, US) according to the manufacturer's instructions, with the following modification: after the addition of 450 µl Lysis Solution MBL during the cell lysis step, the samples were incubated for 10 min in 65 °C and for additional 10 min in 95 °C. The extracted DNA was stored at −20 °C until processing. Flowchart of the process of bacterial DNA isolation and characterization is presented in Figure 2.

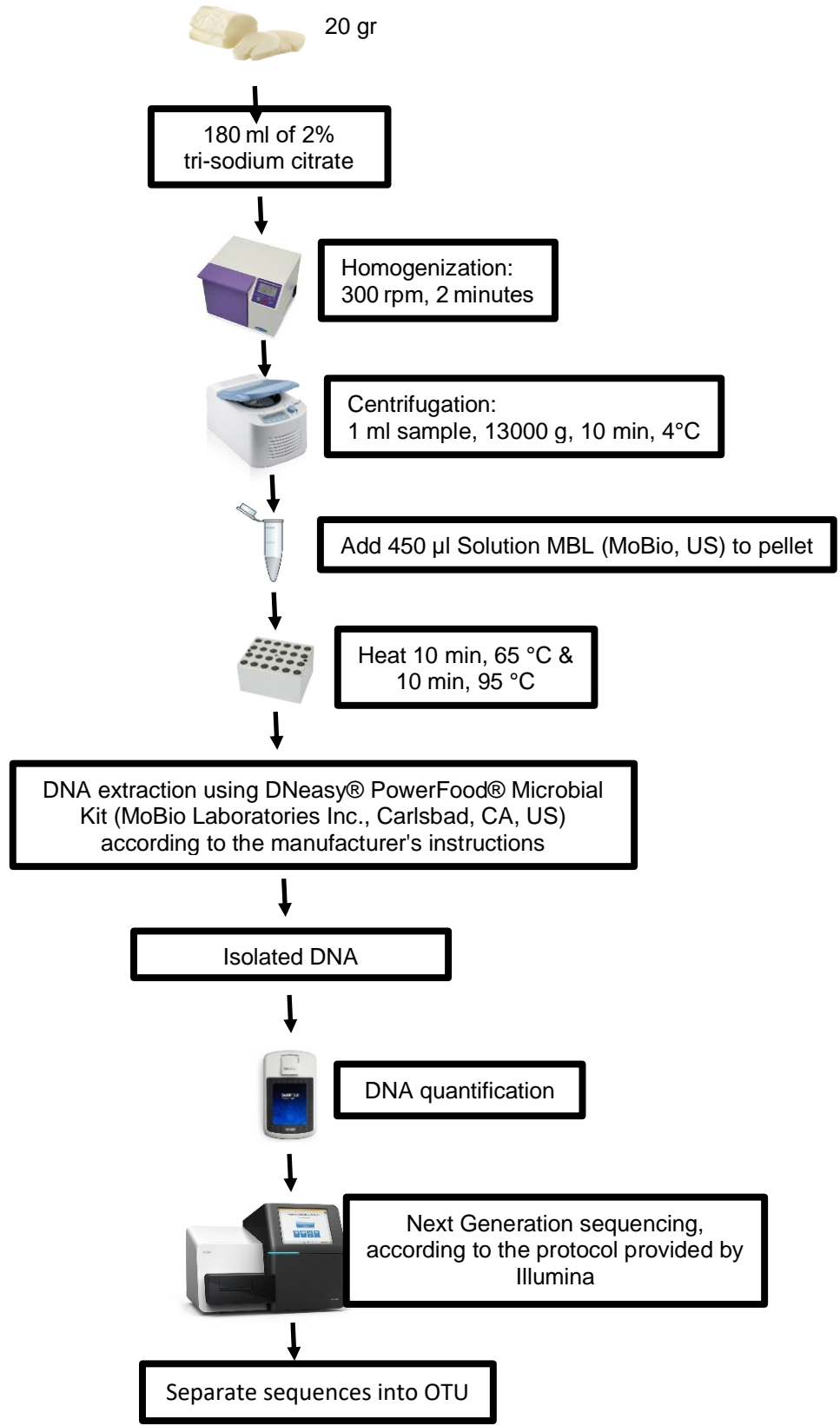

**Figure 2.** Flowchart of bacteria isolation and characterization process.

**Table 1**. Information regarding the Halloumi samples used in the present study.

| Product no | Area of production | Type of milk | Description |
|---|---|---|---|
| G_1 | Pachna, Limassol | Goat | Traditional |
| G_2 | Anarita, Paphos | Goat | Traditional - Mature |
| S_1 | Akaki, Nicosia | Sheep | Traditional |
| GS_1 | Anogyra - Limassol | Goat & sheep | Traditional |
| GS_2 | Akaki, Nicosia | Goat & sheep | Traditional |
| GS_3 | Strovolos, Nicosia | Goat & sheep | Traditional |
| GS_4 | Paramitha, Limassol | Goat & sheep | Traditional |
| GS_5 | Aradipou, Larnaca | Goat & sheep | Traditional |
| GS_6 | Strovolos, Nicosia | Goat & sheep | Traditional |
| GS_7 | Simou, Paphos | Goat & sheep | Traditional |
| GS_8 | Anagia, Nicosia | Goat & sheep | Traditional - mature |
| GS_9 | Athienou, Larnaca | Goat & sheep | Traditional |
| GSC_1 | Ypsonas, Limassol | Goat & sheep & cow | Industrial |
| GSC_2 | Dasaki Achnas, Famagusta | Goat & sheep & cow | Industrial |
| GSC_3 | Strovolos, Nicosia | Goat & sheep & cow | Industrial |
| GSC _4 | Aradipou, Larnaca | Goat & sheep & cow | Industrial |
| GSC _5 | Aradipou, Larnaca | Goat & sheep & cow | Industrial |
| GSC _6 | Akaki, Nicosia | Goat & sheep & cow | Industrial |

### *2.3 Quantification of total DNA*

The total DNA isolated from the Halloumi samples was quantified fluorometrically with Qubit 3.0 fluorometer (Invitrogen, Carlsbad, CA) using Qubit dsDNA HS Assay Kit (Invitrogen). The purity of the DNA was evaluated by measuring the ratio of absorbance A260/280 nm and A260/230 nm using spectrophotometer (NanoDrop Thermo Scientific, USA).

### *2.4 Barcoded Illumina MiSeq amplicon sequencing of bacterial 16s rRNA gene*

The amplification of 16S rRNA bacterial gene - was performed using primers targeting the V3–V4 hyper-variable region using the paired-end approach according to the protocol provided by Illumina (https://support.illumina.com/documents/documentation/chemistry_documentation/16s/16s-metagenomic-library-prep-guide-15044223-b.pdf). The 16S rDNA V3–V4 amplicon was amplified using KAPA HiFi Hot Start Ready Mix (2X) (TaKaRa Bio Inc., Japan). The two universal bacterial 16S rRNA gene amplicon PCR primers used were forward primer (TCGTCGGCAGCGTCAGATGTGTATAAGAGACAG) and reverse primer (GTCTCGTGGGCTCGGAGATGTGTATAAGAGACAG) with the addition of the overhang adapter sequence. Nextera XT Index Kit (FC-131-2001, FC-131-2002) was used for the multiplexing step. The DNA concentration of each PCR product was determined using Qubit dsDNA High sensitivity assay and quality was assessed using a bioanalyzer (Agilent 2200 TapeStation) (expected size

~550 bp). All amplicon products from different samples were mixed in equal concentrations and purified using Agencourt Ampure Beads (Agencourt Bioscience Corporation, MA, USA). The sequencing runs were performed using a MiSeq 300 cycle Reagent Kit v2 (Illumina, USA), on a MiSeq Illumina sequencing platform.

### *2.5 Microbiome and Statistical analysis*

For 16S rRNA DNA sequence clustering and Operational Taxonomic Unit (OTU) filtering, Ribosomal Database Project Classifier (Wang, Garrity, Tiedje, & Cole, 2007) against the Illumina-curated version of GreenGenes reference taxonomy database (DeSantis et al., 2006) was used. The classified organisms were converted to percentages in each sample in order to understand the representation of each organism in the sample and OTU representing less than 0.001% were excluded. Alpha diversity metrics (Shannon, Inverse Simpson and Chao1) were estimated using the EstimateS version 9.1.0 for Windows 10 (http://viceroy.eeb.uconn.edu/estimates/index.html). For Inverse Simpson diversity significance, as well as among bacterial species significant differences, between the traditional and the industrial groups, Mann-Whitney U test (MACKLIN, 1947) was applied, using SPSS statistical package (v.18.0 for Windows; SPSS, Chicago, IL, USA). Stacked Column chart and Weighted Venn diagram of the major OTUs in genus level filtered at 5% and 1% abundance respectively in at least one sample were used for the co-occurrence/co-exclusion analysis using Microsoft Excel and Microsoft Word respectively. Heatmap of hierarchical clustering of the dominant bacterial genera per Halloumi sample was performed using the Seaborn library of Python (https://seaborn.pydata.org).

All sequence data were deposited in Sequence Read Archive (SRA) under BioProject PRJNA598815.

### 3. Results

The present study was performed to characterize the bacterial communities present in Cyprus Halloumi samples, based on the mixture of milk used and the area of production, considering that the manufacturing conditions were according to the standards, using the HTS approach.

### *3.1 Abundance and diversity of members of the bacterial microbiota*

Eighteen (18) examined sample sets were used as input to the Illumina MiSeq in order to generate:

a) 2,791,099 high quality sequencing reads, with an average of 155,061.1 sequencing reads per sample (range = 16,621–630,104, STD = 16,650.4) at the family level,
b) 2,764,127 high quality sequencing reads, with an average of 153,562.6 sequencing reads per sample (range = 16,485–625,388, STD = 16,5385.4) at the genus level and
c) 2,411,369 high quality sequencing reads, with an average of 133,964.94 sequencing reads per sample (range = 12,541– 577,516, STD = 155,939.98) at the species level, and average length of 548 bp (Table S1). High quality sequences were grouped into average number 234.55 OTUs (range = 142- 435, SD = 91.17). Shannon, inverse Simpson and Chao1 estimators for genus level are also shown in Table S1.

Higher diversity was observed in samples produced with the "traditional" method, made by goat and/or sheep milk, in comparison to halloumi made by a mixture of goat, sheep and cow milk, according to the "industrial" method, as was indicated with Inverse Simpson estimator (Figure 3). The two diverse manufacturing conditions differed significantly (p≤0.05) based on Mann-Whitney U test.

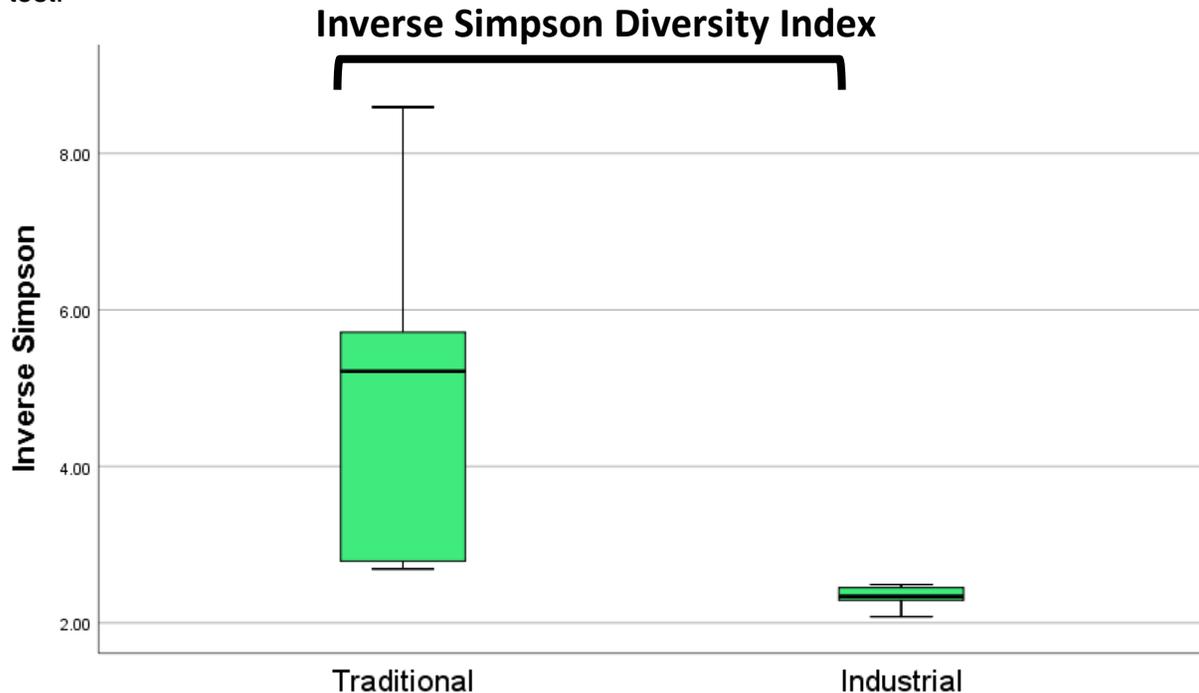

**Figure 3**. Boxplots showing alpha diversity analysis metrics with Inverse Simpson difference between traditionally and industrially made halloumi. *p=0.000 from the Mann-Whitney U test.

### *3.2 Taxonomic composition of bacterial communities in Halloumi samples*

Using 16S rRNA gene sequencing, seven bacterial phyla were identified, consisting mostly of Firmicutes, and Proteobacteria, as well as Actinobacteria and Bacteroidetes, along with three additional phyla (Chloroflexi, Cyanobacteria and Tenericutes). Analysis of the relative abundance of bacterial phyla revealed that Firmicutes and Proteobacteria were the predominant phyla in all Halloumi samples tested. This finding is consistent with other cheese-related 16S rDNA metagenomic studies (Dalmasso et al., 2016; Giello et al., 2017).

The predominant bacteria comprised LAB genera, such as *Lactobacillus, Leuconostoc*, *Pediococcus, Weissella* and the halophilic, alkaliphilic *Marinilactibacillus* (Figure 4). Additionally, spore forming bacteria, including the genus *Bacillus,* psychrophilic or psychrotolerant bacterial genera, such as *Psychrobacter,* the halophilic genus *Halomonas*, as well as the genera *Pseudomonas*, *Staphylococcus*, *Acinetobacter*, *Macrococcus* and *Vibrio*, member of which may cause food spoilage (Li et al., 2014; F. Liu, Wang, Du, Zhu, & Xu, 2010; Towner, 1992) were also commonly detected. The most commonly detected species included the members of the genera *Lactobacillus*, such as *L. manihotivorans*, *L. alimentarius*, *L. brevis* and *L. parakefiri* and *Marinilactibacillus,* such as *M. psychrotolerans*.

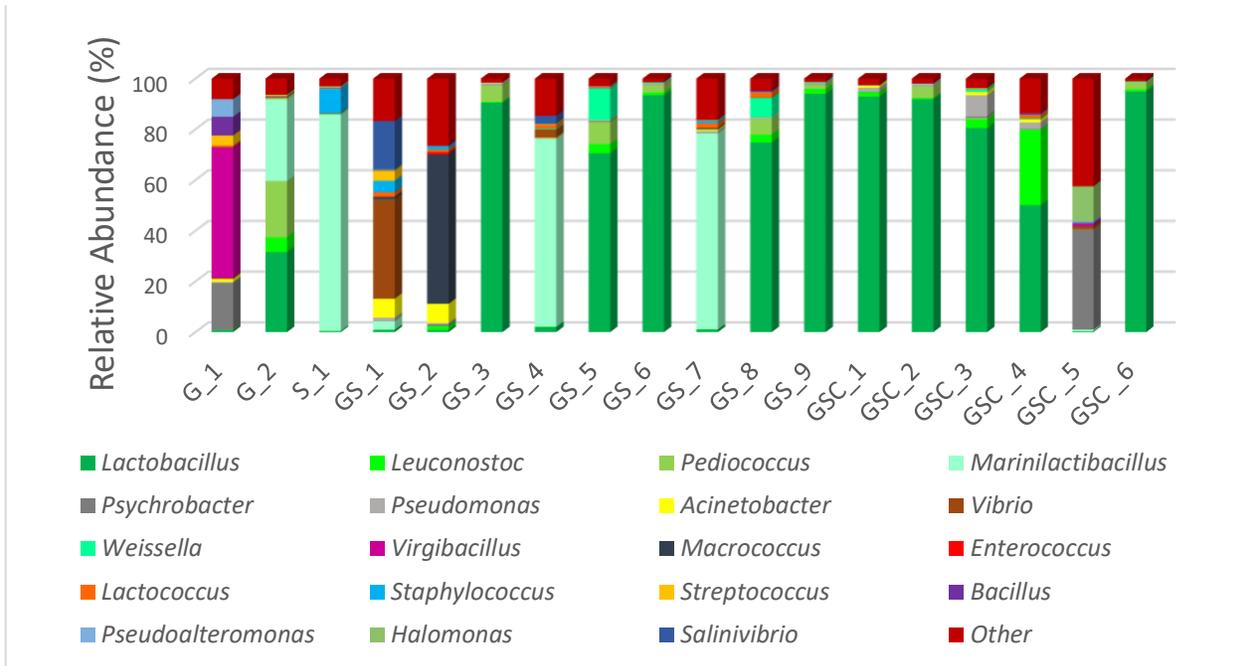

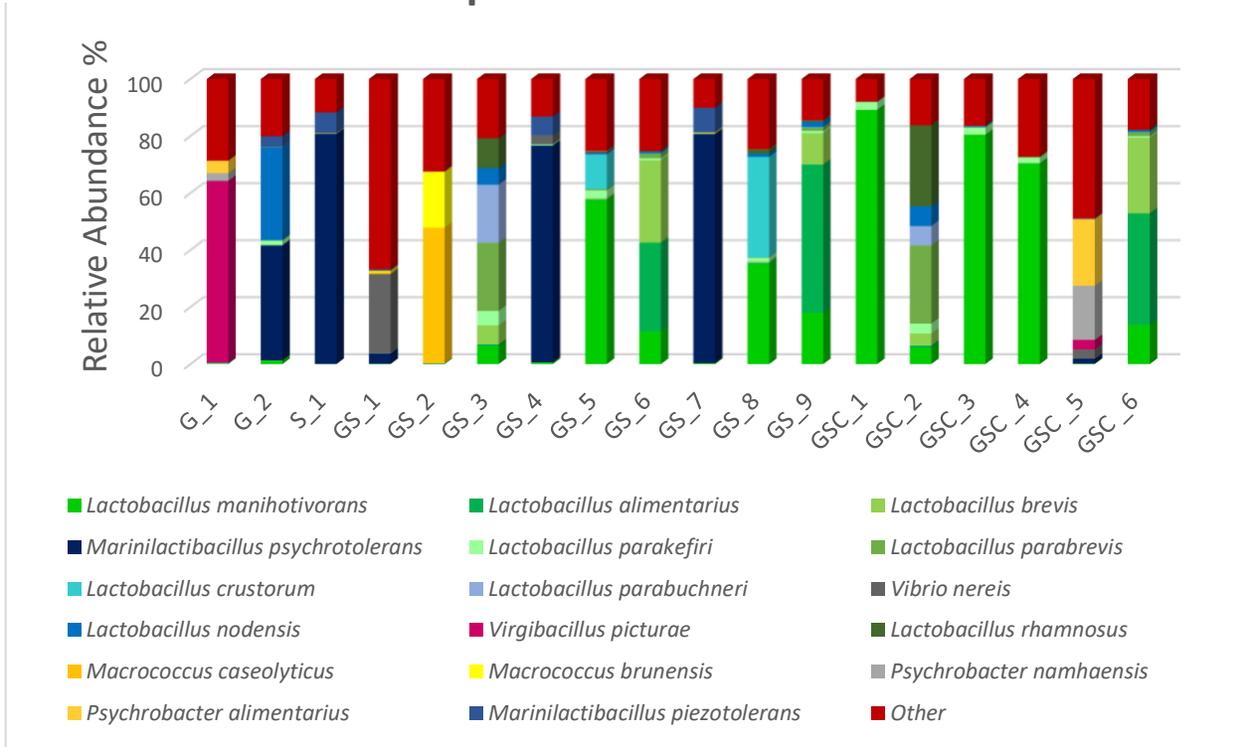

**Figure 4**. 3D 100% Stacked Column chart of the relative abundance of the major taxonomic groups detected by HTS at genus (A), and species (B) level for 18 Halloumi samples. Only OTUs with an incidence above 5% in at least one sample are shown.

Halloumi samples produced according to the "traditional" method (without the addition of cow milk) were distinguished by increased variability in their bacterial communities among samples compared to those produced according to the "industrial" method (with the addition of cow milk) (Figures 3-5). The industrial group was characterized by increased relative abundance of LAB, including the genus *Lactobacillus* (50% - 95%), as well as the genera *Leuconostoc* (0% – 30%) and *Pediococcus* (0% – 4.5%), with the exception of the sample GSC_5. In contrast, in the "traditional" Halloumi group, apart from the genera *Lactobacillus* (0.5% - 94%), *Leuconostoc* (0% – 3.5%) and *Pediococcus* (0% – 9%), additional genera, including *Marinilactibacillus* (0% - 85%), *Staphylococcus* (0% - 9%), *Streptococcus* (0 - 4%), *Macrococcus* (0% – 59%), and *Bacillus* (0% – 7%), dominated the microbiome in some samples. Additional commonly detected genera in both production methods include *Psychrobacter, Pseudomonas, Acinetobacter, Vibrio, Weissella, Virgibacillus* and *Lactococcus* (Figure 5). However, statistical comparison of the two manufacturing conditions indicated no significant differences between the memberships (both in genus and species level) of the samples. The species that mostly indicated difference in relative abundance was *L. manihotivorans* (p=0.05) (Figure S1 A). Noteworthy, samples G_2 and GS_8, which were ripened for 40 days, were also characterized by increased relative abundance of LAB, including the genera *Lactobacillus* (31.5%, 75% respectively), *Leuconostoc* (6%, 3% respectively), *Pediococcus* (22%, 6% respectively), *Marinilactibacillus* (32%, 0% respectively) and *Weissella* (0%, 8% respectively).

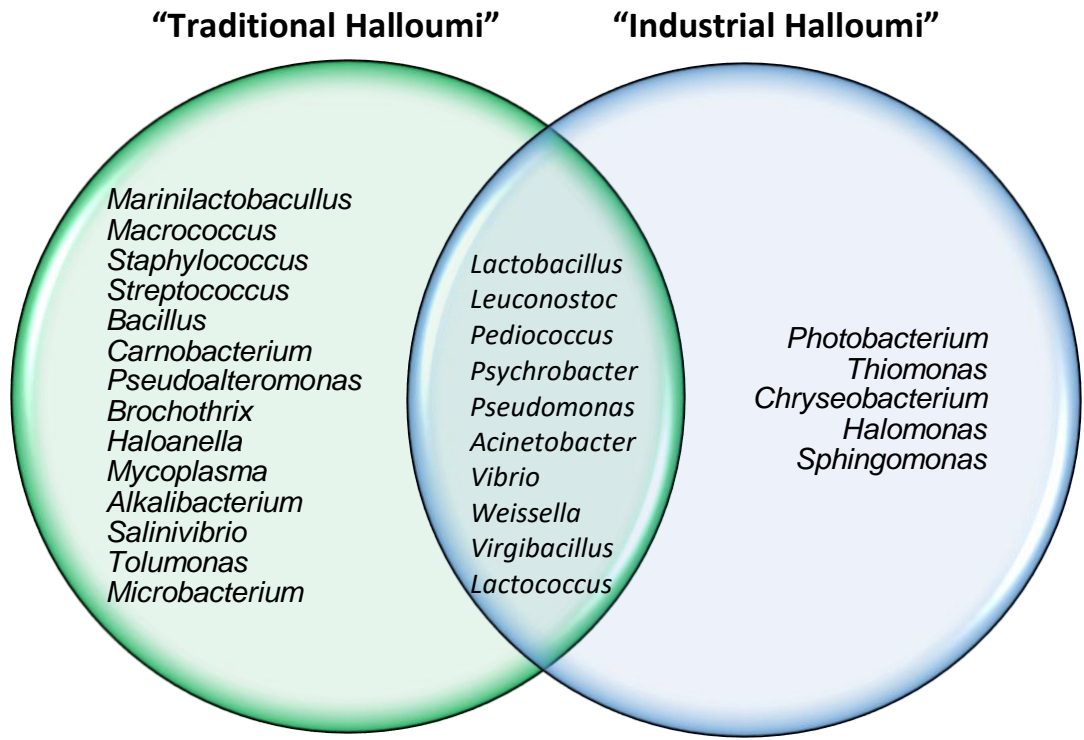

**Figure 5**. Weighted Venn diagram representing shared and unique bacterial compositions in "traditional" and "industrial" Halloumi. The bacterial OTUs at the genus levels were defined with 16S rRNA gene sequences based on a cut-off value of >97% sequence similarity. Only OTUs with an incidence above 1% in at least one sample are shown.

## 3.3 Relationships between Bacterial Communities among Halloumi Samples

To get an overall view on the identified associations among Halloumi samples, based on their bacterial communities' structure, a hierarchically clustered heatmap was produced (Figure 6). The heatmap plot depicted the normalized relative abundance of the dominant bacterial genera (relative abundance >1% in two samples or >3% in one sample) (variables clustering on the Y-axis) within each sample (X-axis clustering). The hierarchically clustered heatmap revealed that:

1) Most of Halloumi samples were clustered together. This cluster was characterized by elevated relative abundance of the genus *Lactobacillus*. Other genera also associated with *Lactobacillus* were *Pediococcus* and *Weissella*, as well as *Leuconostoc* and *Pseudomonas* in a lower degree. This cluster included all the industrial samples but sample GSC_5. Additionally, samples GS_3, GS_5, GS_6 and GS_9 were also included.
2) Sample GSC_5 was related to sample G_1 in a separate cluster, characterized by the presence of *Psychrobacter.*
3) Sample GS_2 was differentiated from the other samples by the presence of *Macrococcus.*
4) Another cluster was characterized by the presence of *Marinilactobacillus*. This cluster included samples S_1, GS_4 and GS_7.
5) Finally, GS_1 created a separate cluster close related to the elevated relative abundance of *Marinilactobacillus* cluster, and was characterized by elevated relative abundance of *Salinivibrio* and *Vibrio*.

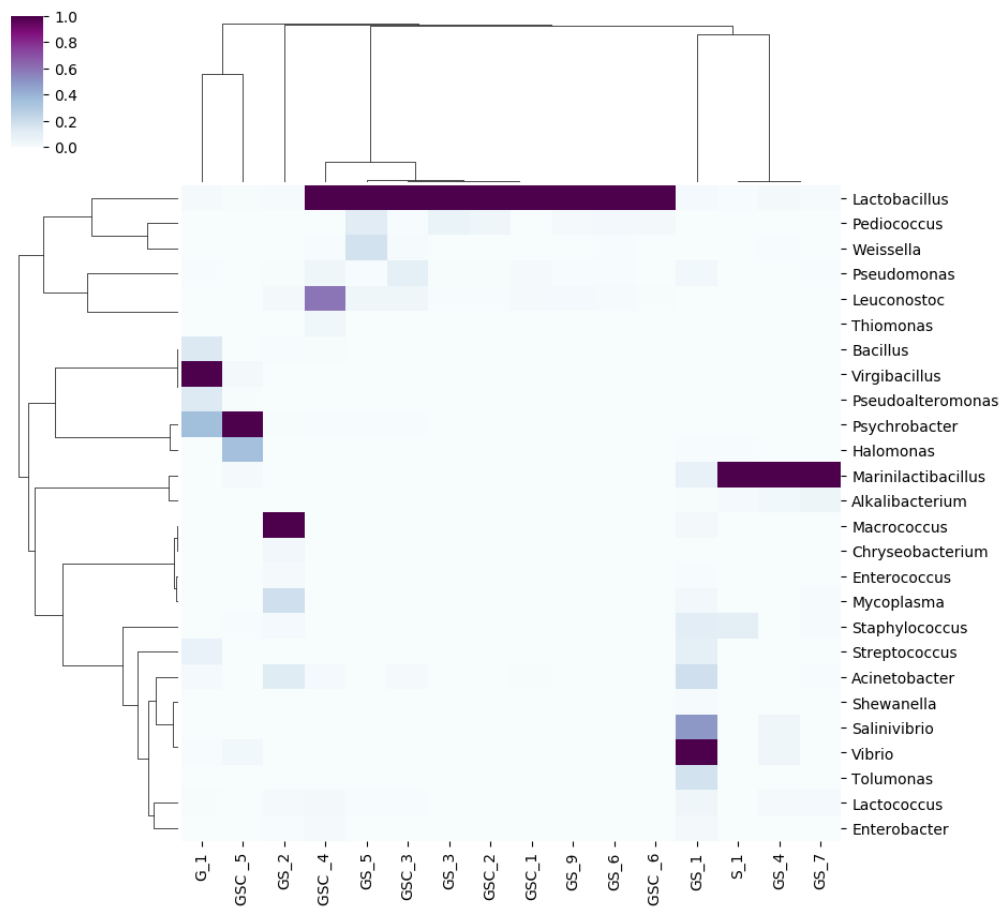

**Figure 6**. Heatmap of hierarchical clustering of the dominant bacterial genera (% relative sequence abundance ≥ 1 in at least two samples or ≥ 3 in one sample) represented by 16S ribosomal RNA (rRNA) amplicons per halloumi sample. Heatmap color (white to green) displays the row scaled relative abundance of each taxon across all samples. Heatmap rows and columns distance metric were clustered based on Pearson correlation.

## 4. Discussion

The present study is the first -to our knowledge- which targets to characterize the bacterial communities established in Cyprus Halloumi by metagenomic analysis. HTS was applied for an in-depth quantitative characterization of the structure of the bacterial population. The study aimed at identifying how the microbiota is shaped in Cyprus Halloumi and whether the different milk mixtures used for Halloumi manufacturing affect the bacterial communities' formation of the final product. Based on the fact that (according to the standard CYS 94) Halloumi cheese is made from fresh pasteurized milk, the microbiota was expected to consist of thermophilic bacteria, in accordance to 16S rDNA metagenomic studies performed in other brined cheeses made by pasteurized milk, such as the PDO Oscypek cheese (Alegría, Szczesny, Mayo, Bardowski, & Kowalczyk, 2012) and Gouda cheese (Salazar et al., 2018), but also in other PDO cheeses including Herve (Delcenserie et al., 2014) and Pecorino Toscano cheese (De Pasquale et al., 2016). OTU analysis of the data from 16S rRNA gene sequences revealed that the bacterial communities of Halloumi samples in their majority were dominated by species of the genus *Lactobacillus.* Their capacity to survive in low pH and high salt concentrations in combination to successfully ferment lactose makes them predominant in several cheeses, as well as contributors to sensorial characteristics development (Fernández et al., 2015, Kamilari et al., 2019). Additional genera that co-occurred in some Halloumi samples, included *Leuconostoc* and *Pediococcus.* The presence of these genera were also reported in various metagenomic studies, including Oscypek cheese (Alegría et al., 2012) and Italian high-moisture Mozzarella cheese (Marino et al., 2019). Remarkably, samples produced according to the "industrial" method were characterized by significantly lower diversity in genus level, compared to Halloumi manufactured according to the "traditional" method. However, the lower bacterial diversity could not be attributed to the addition of cow milk. A recent study, indicated that Halloumi made by 100% goat's milk had almost the same to slightly higher total bacterial counts compared to Halloumi made by different mixtures of cow and goat milk (Elgaml, Moussa, & Saleh, 2017). Quigley et al. (Quigley et al., 2012) indicated that Irish artisanal cheeses made by cow milk had different and richer bacterial diversity compared to cheeses made by goat or sheep milk. Based on this, the reduction in the bacterial diversity in "industrial" Halloumi, is most probably associated with the manufacturing conditions applied (i.e. harsher heat treatments).

Some "traditionally" produced Halloumi were characterized by the dominant presence of the halophilic LAB *Marinilactibacillus* (Figure 4). Species which belong to this genus might have been transferred to Halloumi through sea salt originated from the Cyprus marine environment, in accordance with a similar report by Yumoto et al. (Yumoto, Hirota, & Yoshimune, 2011). The 16S rDNA metagenomic analysis study of Halitzia cheese, another white-brined cheese of Cyprus, reported the dominant presence of *Lactobacillus* and *Leuconostoc*, in addition to *Lactococcus,* but not of *Marinilactibacillus* (Papademas et al., 2019).

The most commonly detected species included the members of the genera *Lactobacillus*, such as *L. manihotivorans*, *L. alimentarius*, *L. brevis* and *L. parakefiri* and *Marinilactibacillus,* including *M. psychrotolerans*. The existence of these species in cheese samples is familiar from other studies (Alegría et al., 2012; Aquilanti et al., 2011; Bora & Ward, 2015; Ishikawa et al., 2013; Lacerda et al., 2011; Ledina et al., 2018). In comparison with other white brined cheeses, *L. brevis* was detected in the traditional Greek Feta PDO cheese (Bozoudi et al., 2016; Rantsiou, Urso, Dolci, Comi, & Cocolin, 2008) and the Spanish PDO Alberquilla cheese (Abriouel, Martín-Platero, Maqueda, Valdivia, & Martínez-Bueno, 2008), and *Leuconostoc mesenteroides* in Feta (Bozoudi et al., 2016) and the traditional Egyptian soft Domiati cheese (El-Baradei, Delacroix-Buchet, & Ogier, 2007), via PCR-denaturing gradient gel electrophoresis (DGGE) analysis.

The dominant species of Halloumi differed from those found via metagenomic studies performed in the white brined cheeses Mozzarella (PDO) and Parmigiano Reggiano (PDO), in that *Streptococcus thermophilus*, *Lactococcus lactis, Lactobacillus delbrueckii, Lactobacillus helveticus, Lactobacillus kefiranofaciens,* as well as *Lactobacillus fermentum,* and *Lactobacillus delbrueckii*, *Lactobacillus helveticus* and *S. thermophilus* respectively, were the most abundant bacterial species (De Filippis, La Storia, Stellato, Gatti, & Ercolini, 2014; Ercolini, De Filippis, La Storia, & Iacono, 2012; Marino et al., 2019). However, the microbiome of these cheeses is affected by the addition of natural whey cultures (NWCs), in which thermophilic and mesophilic LAB are the most dominant. In addition, HTS methodologies indicated that the microbiome of the traditional feta like Iranian cheese Liqvan was characterized by different LAB species compared to Halloumi, such as *Lactobacillus curvatus*, *Lactobacillus zeae*, *Lactobacillus fuchunensis*, *Lactococcus lactis* and *Lactobacillus pentosus* and *Lactobacillus kefiri* (Ramezani, Hosseini, Ferrocino, Amoozegar, & Cocolin, 2017)*.* The increased representation of LAB in Halloumi, including *Lactobacillus* sp., *Leuconostoc* sp. *Pediococcus* sp. and *Weissella* sp. members of which possess antimicrobial properties, reduces the risk of colonization of food-borne diseases causing bacteria, such as *Listeria monocytogenes*, *Staphylococcus aureus*, *Salmonella* spp., and pathogenic *Escherichia coli* (Arqués, Rodríguez, Langa, Landete, & Medina, 2015) providing an indication of improved safety of the product for consumers. Interestingly, the species *Marinilactibacillus psychrotolerans* is also known to possess an anti-Listeria effect (Roth, Schwenninger, Eugster-Meier, & Lacroix, 2011).

Members of the family Bacillaceae, including *Bacillus* and *Virgibacillus* were also detected in increased relative abundance in one traditional Halloumi made by goat milk. The dominant occurrence of this family in addition to *Lactococcus* was reported in Gouda cheese (Salazar et al., 2018). Species of this family are considered contaminants from milk or might have originated from other sources (e.g. dry *Mentha viridis* leaves) and their capability to create biofilms as well as heat-resistant endospores, allow them to survive from sanitization processes. Elevated relative abundance of members of the family Vibrionaceae, such as *Salinivibrio* and *Vibrio* was additionally found in sample GS_1. These salt-tolerant contaminants might have originated also from the marine environment (Sawabe et al., 2013). Further contaminants that were detected in increased relative abundance in samples G_1 and GSC _5, include the genera *Pseudoalteromonas* and *Halomonas*, and *Psychrobacter*. The presence of these genera has also been detected in Herve (PDO) cheese (Delcenserie et al., 2014) and could be explained by preservation in brine before packaging. The existence of salt-tolerant bacteria and their adaptation to a high-salt environment has been reported in metagenomic studies of artisanal cheeses (Fuka et al., 2013; Quigley et al., 2012).

Furthermore, three Halloumi samples produced according to the traditional method, were characterized by an increased representation of the genera *Acinetobacter* and *Macrococcus* and/or *Streptococcus*. These contaminants have been commonly detected in raw milk (Liu *et al.*, 2015; Seon Kim *et al.*, 2017) and cheeses, including Poro (Aldrete-Tapia, Escobar-Ramírez, Tamplin, & Hernández-Iturriaga, 2014), Danish (Masoud et al., 2011) and Pico cheese (Riquelme et al., 2015). The two *Macrococcus* species detected, *M. caseolyticus* and *M. brunensis* are reported as able to breakdown casein, contributing to aroma precursors formation (Fuka et al., 2013). Finally, one sample produced according to the "traditional" way had elevated relative abundance of *Mycoplasma,* species of which may cause infections in sheep and goats (Woubit et al., 2007). Their existence in Halloumi may again be associated with improper pasteurization process, environmental contamination and unsuccessful storage conditions. Indicatively, analysis of Grana-like hard cheeses, showed that the use of natural whey cultures rich in LAB during cheesemaking, did not suffice to prevent the development of spoilage bacteria in mature cheese, when raw cow's milk rich in contaminants and spoilage bacteria was used during manufacturing (Alessandria et al., 2016)

Noteworthy, the analysis included two mature Halloumi samples (G_2 and GS_8). Both samples were characterized by elevated relative abundance of LAB, such as *Lactobacillus*, *Leuconostoc*, *Pediococcus*, *Marinilactibacillus* and *Weissella*. Increased representation of *Lactobacillus plantarum*, *Leuconostoc mesenteroides* and *Weissella paramesenteroides* was also indicated in the long ripened traditional Mexican Cotija cheese (Escobar-Zepeda, Sanchez-Flores, & Quirasco Baruch, 2016). The process of ripening was additionally found to promote the dominance of LAB in three traditional Croatian ewe's milk cheeses, called Istrian, Krcki and Paski (Fuka et al., 2013). The increased proteolytic and lipolytic capacity as well as the ability of LAB bacteria to metabolize additional carbon sources, provides them the advantage to predominate the microenvironment of ripened cheeses (Kamilari et al., 2019). The degradation of those compounds during ripening, releases volatile organic compounds, contributing to different sensorial characteristics' development in mature, compare to fresh halloumi samples (Papademas & Robinson, 2000).

## 5. Conclusion

This is the first study performed to characterize the bacterial communities of Cyprus Halloumi cheese via HTS. The study highlights the significant influence of the manufacturing method i.e. "traditional" versus "industrial", in respect to Halloumi's established microbiome. The harsher heat treatments performed by industries could shape microbial communities. Additionally, the study proves that the microbiome could be used as a possible fingerprint for the characterization of the typicity of Cyprus Halloumi cheese. In the future, additional samples are to be analyzed in order to compare the microbiome of cheeses similar to Halloumi cheese possibly made in other countries. HTS analysis could be combined with additional methodologies, such as major and trace elements quantification via inductively coupled plasma–atomic emission spectroscopy (ICP-AES), an analysis that indicated significant differences in Halloumi cheese samples produced in different geographical areas (Osorio, Koidis, & Papademas, 2015), or isotopic analysis, in order to offer a more thorough characterization of the fingerprint of Cyprus Halloumi.

**Acknowledgments**

The authors acknowledge the funding from the project AGRO-ID, INTERREG Greece-Cyprus 2014-2020.

Andreas Kamilaris has received funding from the European Union's Horizon 2020 Research and Innovation Programme under grant agreement No 739578 complemented by the Government of the Republic of Cyprus through the Directorate General for European Programmes, Coordination and Development.

The authors declare no conflict of interest.

**References**

Abriouel, H., Martín-Platero, A., Maqueda, M., Valdivia, E., & Martínez-Bueno, M. (2008). Biodiversity of the microbial community in a Spanish farmhouse cheese as revealed by culture-dependent and culture-independent methods. *International Journal of Food Microbiology*. https://doi.org/10.1016/j.ijfoodmicro.2008.07.004

Aldrete-Tapia, A., Escobar-Ramírez, M. C., Tamplin, M. L., & Hernández-Iturriaga, M. (2014). High-throughput sequencing of microbial communities in Poro cheese, an artisanal Mexican cheese. *Food Microbiology*. https://doi.org/10.1016/j.fm.2014.05.022

Alegría, Á., Szczesny, P., Mayo, B., Bardowski, J., & Kowalczyk, M. (2012). Biodiversity in Oscypek, a traditional Polish Cheese, determined by culture-dependent and -independent approaches. *Applied and Environmental Microbiology*. https://doi.org/10.1128/AEM.06081-11

Alessandria, V., Ferrocino, I., De Filippis, F., Fontana, M., Rantsiou, K., Ercolini, D., & Cocolin, L. (2016). Microbiota of an Italian Grana-like cheese during manufacture and ripening, unraveled by 16S rRNA-based approaches. *Applied and Environmental Microbiology*. https://doi.org/10.1128/AEM.00999-16

Aquilanti, L., Babini, V., Santarelli, S., Osimani, A., Petruzzelli, A., & Clementi, F. (2011). Bacterial dynamics in a raw cow's milk Caciotta cheese manufactured with aqueous extract of Cynara cardunculus dried flowers. *Letters in Applied Microbiology*. https://doi.org/10.1111/j.1472-765X.2011.03053.x

Arqués, J. L., Rodríguez, E., Langa, S., Landete, J. M., & Medina, M. (2015). Antimicrobial activity of lactic acid bacteria in dairy products and gut: Effect on pathogens. *BioMed Research International*. https://doi.org/10.1155/2015/584183

Bintsis, T., & Papademas, P. (2002). Microbiological quality of white-brined cheeses: A review. *International Journal of Dairy Technology*. https://doi.org/10.1046/j.1471-0307.2002.00054.x

Bokulich, N. A., & Mills, D. A. (2013). Facility-specific "house" microbiome drives microbial landscapes of artisan cheesemaking plants. *Applied and Environmental Microbiology*. https://doi.org/10.1128/AEM.00934-13

Bora, N., & Ward, A. C. (2015). Analyzing the metagenome of smear cheese flora using next generation sequencing tools. In *Diversity, Dynamics and Functional Role of Actinomycetes on European Smear Ripened Cheeses*. https://doi.org/10.1007/978-3-319-10464-5_5

Bozoudi, D., Torriani, S., Zdragas, A., & Litopoulou-Tzanetaki, E. (2016). Assessment of microbial diversity of the dominant microbiota in fresh and mature PDO Feta cheese made at three mountainous areas of Greece. *LWT - Food Science and Technology*. https://doi.org/10.1016/j.lwt.2016.04.039

Calasso, M., Ercolini, D., Mancini, L., Stellato, G., Minervini, F., Di Cagno, R., … Gobbetti, M. (2016). Relationships among house, rind and core microbiotas during manufacture of traditional Italian cheeses at the same dairy plant. *Food Microbiology*. https://doi.org/10.1016/j.fm.2015.10.008

CYS 94 (1985) Standard for fresh Halloumi cheese. Cyprus Organization for Standards and Control of Quality 1-3; Nicosia

Dalmasso, A., Soto del Rio, M. de los D., Civera, T., Pattono, D., Cardazzo, B., & Bottero, M. T. (2016). Characterization of microbiota in Plaisentif cheese by high-throughput sequencing. *LWT - Food Science and Technology*. https://doi.org/10.1016/j.lwt.2016.02.004

De Filippis, F., La Storia, A., Stellato, G., Gatti, M., & Ercolini, D. (2014). A selected core microbiome drives the early stages of three popular Italian cheese manufactures. *PLoS ONE*. https://doi.org/10.1371/journal.pone.0089680

De Pasquale, I., Di CagnWolfe, B. E. et al. (2014) 'Cheese rind communities provide tractable systems for in situ and in vitro studies of microbial diversity', Cell. doi: 10.1016/j.cell.2014.05.041.o, R., Buchin, S., De Angelis, M., & Gobbetti, M. (2016). Spatial distribution of the metabolically active microbiota within Italian PDO ewes' milk cheeses. *PLoS ONE*. https://doi.org/10.1371/journal.pone.0153213

Delbès, C., Ali-Mandjee, L., & Montel, M. C. (2007). Monitoring bacterial communities in raw milk and cheese by culture-dependent and -independent 16S rRNA gene-based analyses. *Applied and Environmental Microbiology*. https://doi.org/10.1128/AEM.01716-06

Delcenserie, V., Taminiau, B., Delhalle, L., Nezer, C., Doyen, P., Crevecoeur, S., … Daube, G. (2014). Microbiota characterization of a Belgian protected designation of origin cheese, Herve cheese, using metagenomic analysis. *Journal of Dairy Science*. https://doi.org/10.3168/jds.2014-8225

DeSantis, T. Z., Hugenholtz, P., Larsen, N., Rojas, M., Brodie, E. L., Keller, K., … Andersen, G. L. (2006). Greengenes, a chimera-checked 16S rRNA gene database and workbench compatible with ARB. *Applied and Environmental Microbiology*. https://doi.org/10.1128/AEM.03006-05

El-Baradei, G., Delacroix-Buchet, A., & Ogier, J. C. (2007). Biodiversity of bacterial ecosystems in traditional Egyptian Domiati cheese. *Applied and Environmental Microbiology*. https://doi.org/10.1128/AEM.01667-06

Elgaml, N., Moussa, M. A. M., & Saleh, A. E. (2017). Comparison of the Properties of Halloumi Cheese Made from Goat Milk, Cow Milk and Their Mixture. *Journal of Sustainable Agricultural Sciences*. https://doi.org/10.21608/jsas.2017.1065.1006

Ercolini, D., De Filippis, F., La Storia, A., & Iacono, M. (2012). "Remake" by high-throughput sequencing of the microbiota involved in the production of water buffalo mozzarella cheese. *Applied and Environmental Microbiology*. https://doi.org/10.1128/AEM.02218-12

Escobar-Zepeda, A., Sanchez-Flores, A., & Quirasco Baruch, M. (2016). Metagenomic analysis of a Mexican ripened cheese reveals a unique complex microbiota. *Food Microbiology*.

https://doi.org/10.1016/j.fm.2016.02.004

Fernández, M., Hudson, J. A., Korpela, R., & De Los Reyes-Gavilán, C. G. (2015). Impact on human health of microorganisms present in fermented dairy products: An overview. *BioMed Research International*. https://doi.org/10.1155/2015/412714

Fuka, M. M., Wallisch, S., Engel, M., Welzl, G., Havranek, J., & Schloter, M. (2013). Dynamics of bacterial communities during the ripening process of different Croatian cheese types derived from raw ewe's milk cheeses. *PLoS ONE*. https://doi.org/10.1371/journal.pone.0080734

Giello, M., La Storia, A., Masucci, F., Di Francia, A., Ercolini, D., & Villani, F. (2017). Dynamics of bacterial communities during manufacture and ripening of traditional Caciocavallo of Castelfranco cheese in relation to cows' feeding. *Food Microbiology*. https://doi.org/10.1016/j.fm.2016.11.016

Ishikawa, M., Yamasato, K., Kodama, K., Yasuda, H., Matsuyama, M., Okamoto-Kainuma, A., & Koizumi, Y. (2013). Alkalibacterium gilvum sp. nov., slightly halophilic and alkaliphilic lactic acid bacterium isolated from soft and semi-hard cheeses. *International Journal of Systematic and Evolutionary Microbiology*. https://doi.org/10.1099/ijs.0.042556-0

Kamilari E., Tomazou M., Antoniades A., Tsaltas D. High Throughput Sequencing Technologies as a New Toolbox for Deep Analysis, Characterization and Potentially Authentication of Protection Designation of Origin Cheeses? International Journal of Food Science https://doi.org/10.1155/2019/5837301

Kim, D. S., Choi, S. H., Kim, D. W., Kim, R. N., Nam, S. H., Kang, A., … Park, H. S. (2011). Genome sequence of Lactobacillus cypricasei KCTC 13900. *Journal of Bacteriology*. https://doi.org/10.1128/JB.05659-11

Lacerda, I. C. A., Gomes, F. C. O., Borelli, B. M., Faria, C. L. L., Franco, G. R., Mourão, M. M., … Rosa, C. A. (2011). Identification of the bacterial community responsible for traditional fermentation during sour cassava starch, cachaça and minas cheese production using cultureindependent 16s rrna gene sequence analysis. *Brazilian Journal of Microbiology*. https://doi.org/10.1590/S1517-83822011000200029

Lawson, P. A., Papademas, P., Wacher, C., Falsen, E., Robinson, R., & Collins, M. D. (2001). Lactobacillus cypricasei sp. nov., isolated from Halloumi cheese. *International Journal of Systematic and Evolutionary Microbiology*. https://doi.org/10.1099/00207713-51-1-45

Ledina, T., Mohar-Lorbeg, P., Golob, M., Djordjevic, J., Bogovič-Matijašić, B., & Bulajic, S. (2018). Tetracycline resistance in lactobacilli isolated from Serbian traditional raw milk cheeses. *Journal of Food Science and Technology*. https://doi.org/10.1007/s13197-018-3057-6

Li, K., Lin, K., Li, Z., Zhang, Q., Song, F., Che, Z., … Xiang, W. (2014). Spoilage and pathogenic bacteria associated with spoilage process of Sichuan Pickle during the spontaneous fermentation. *Food Science and Technology Research*. https://doi.org/10.3136/fstr.20.899

Liu, F., Wang, D., Du, L., Zhu, Y., & Xu, W. (2010). Diversity of the predominant spoilage bacteria in water-boiled salted duck during storage. *Journal of Food Science*. https://doi.org/10.1111/j.1750-3841.2010.01644.x

Liu, W., Zheng, Y., Kwok, L. Y., Sun, Z., Zhang, J., Guo, Z., … Zhang, H. (2015). High-throughput sequencing for the detection of the bacterial and fungal diversity in Mongolian naturally fermented cow's milk in Russia. *BMC Microbiology*. https://doi.org/10.1186/s12866-015-0385-9

MACKLIN, M. T. (1947). FALLACIES INHERENT IN THE PROBAND METHOD OF ANALYSIS OF HUMAN PEDIGREES FOR INHERITANCE OF RECESSIVE TRAITS. *American Journal of Diseases of Children*. https://doi.org/10.1001/archpedi.1947.02030010469005

Marino, M., Dubsky de Wittenau, G., Saccà, E., Cattonaro, F., Spadotto, A., Innocente, N., … Marroni, F. (2019). Metagenomic profiles of different types of Italian high-moisture Mozzarella cheese. *Food Microbiology*. https://doi.org/10.1016/j.fm.2018.12.007

Masoud, W., Takamiya, M., Vogensen, F. K., Lillevang, S., Al-Soud, W. A., Sørensen, S. J., & Jakobsen, M. (2011). Characterization of bacterial populations in Danish raw milk cheeses made with different starter cultures by denaturating gradient gel electrophoresis and pyrosequencing. *International Dairy Journal*. https://doi.org/10.1016/j.idairyj.2010.10.007

Naser, S. M., Vancanneyt, M., Hoste, B., Snauwaert, C., & Swings, J. (2006). Lactobacillus cypricasei Lawson et al. 2001 is a later heterotypic synonym of Lactobacillus acidipiscis Tanasupawat et al. 2000. *International Journal of Systematic and Evolutionary Microbiology*. https://doi.org/10.1099/ijs.0.64229-0

Osorio, M. T., Koidis, A., & Papademas, P. (2015). Major and trace elements in milk and Halloumi cheese as markers for authentication of goat feeding regimes and geographical origin. *International Journal of Dairy Technology*. https://doi.org/10.1111/1471-0307.12213

Papademas, P, & Robinson, R. K. (2000). A comparison of the chemical, microbiological and sensory characteristics of bovine and ovine Halloumi cheese. *International Dairy Journal*. https://doi.org/10.1016/S0958-6946(00)00110-2

Papademas, P, Aspri, M., Mariou, M., Dowd, S. E., Kazou, M., & Tsakalidou, E. (2019). Conventional and omics approaches shed light on Halitzia cheese, a long-forgotten white-brined cheese from Cyprus. *International Dairy Journal*. https://doi.org/10.1016/j.idairyj.2019.06.010

Papademas, P, & Robinson, R. K. (1998). Halloumi cheese: The product and its characteristics. *International Journal of Dairy Technology*. https://doi.org/10.1111/j.1471-0307.1998.tb02646.x

Quigley, L., O'Sullivan, O., Beresford, T. P., Ross, R. P., Fitzgerald, G. F., & Cotter, P. D. (2012). High-throughput sequencing for detection of subpopulations of bacteria not previously associated with artisanal cheeses. *Applied and Environmental Microbiology*. https://doi.org/10.1128/AEM.00918-12

Ramezani, M., Hosseini, S. M., Ferrocino, I., Amoozegar, M. A., & Cocolin, L. (2017). Molecular investigation of bacterial communities during the manufacturing and ripening of semi-hard Iranian Liqvan cheese. *Food Microbiology*. https://doi.org/10.1016/j.fm.2017.03.019

Rantsiou, K., Urso, R., Dolci, P., Comi, G., & Cocolin, L. (2008). Microflora of Feta cheese from four Greek manufacturers. *International Journal of Food Microbiology*. https://doi.org/10.1016/j.ijfoodmicro.2008.04.031

Riquelme, C., Câmara, S., Enes Dapkevicius, M. de L. N., Vinuesa, P., da Silva, C. C. G., Malcata, F. X., & Rego, O. A. (2015). Characterization of the bacterial biodiversity in Pico cheese (an artisanal Azorean food). *International Journal of Food Microbiology*. https://doi.org/10.1016/j.ijfoodmicro.2014.09.031

Roth, E., Schwenninger, S. M., Eugster-Meier, E., & Lacroix, C. (2011). Facultative anaerobic halophilic and alkaliphilic bacteria isolated from a natural smear ecosystem inhibit Listeria growth in early ripening stages. *International Journal of Food Microbiology*. https://doi.org/10.1016/j.ijfoodmicro.2011.02.032

Salazar, J. K., Carstens, C. K., Ramachandran, P., Shazer, A. G., Narula, S. S., Reed, E., … Schill, K. M. (2018). Metagenomics of pasteurized and unpasteurized gouda cheese using targeted 16S rDNA sequencing. *BMC Microbiology*. https://doi.org/10.1186/s12866-018-1323-4

Sawabe, T., Ogura, Y., Matsumura, Y., Feng, G., Rohul Amin, A. K. M., Mino, S., … Hayashi, T. (2013). Updating the Vibrio clades defined by multilocus sequence phylogeny: Proposal of eight new clades, and the description of Vibrio tritonius sp. nov. *Frontiers in Microbiology*. https://doi.org/10.3389/fmicb.2013.00414

Seon Kim, I., Kyung Hur, Y., Ji Kim, E., Ahn, Y. T., Geun Kim, J., Choi, Y. J., & Sung Huh, C. (2017). Comparative analysis of the microbial communities in raw milk produced in different regions of Korea. *Asian-Australasian Journal of Animal Sciences*. https://doi.org/10.5713/ajas.17.0689

Stellato, G., De Filippis, F., La Storia, A., & Ercolini, D. (2015). Coexistence of lactic acid bacteria and potential spoilage microbiota in a dairy processing environment. *Applied and Environmental Microbiology*. https://doi.org/10.1128/AEM.02294-15

Sun, Z., Harris, H. M. B., McCann, A., Guo, C., Argimón, S., Zhang, W., … O'Toole, P. W. (2015). Expanding the biotechnology potential of lactobacilli through comparative genomics of 213 strains and associated genera. *Nature Communications*. https://doi.org/10.1038/ncomms9322

Tamime, A. Y. (2007). *Brined Cheeses. Brined Cheeses*. https://doi.org/10.1002/9780470995860

Towner, K. J. (1992). The Genus Acinetobacter. In *The Prokaryotes*. https://doi.org/10.1007/978-1-4757-2191-1_2

Wang, Q., Garrity, G. M., Tiedje, J. M., & Cole, J. R. (2007). Naïve Bayesian classifier for rapid assignment of rRNA sequences into the new bacterial taxonomy. *Applied and Environmental Microbiology*. https://doi.org/10.1128/AEM.00062-07

Welz G., 2015. European Products: Making and Unmaking Heritage in Cyprus. Berghahn Books. pp. 93–110. ISBN 9781782388234.

Woubit, S., Manso-Silván, L., Lorenzon, S., Gaurivaud, P., Poumarat, F., Pellet, M. P., … Thiaucourt, F. (2007). A PCR for the detection of mycoplasmas belonging to the Mycoplasma mycoides cluster: Application to the diagnosis of contagious agalactia. *Molecular and Cellular Probes*. https://doi.org/10.1016/j.mcp.2007.05.008

Yeluri Jonnala, B. R., McSweeney, P. L. H., Sheehan, J. J., & Cotter, P. D. (2018). Sequencing of the Cheese Microbiome and Its Relevance to Industry. *Frontiers in Microbiology*. https://doi.org/10.3389/fmicb.2018.01020

Yumoto, I., Hirota, K., & Yoshimune, K. (2011). Environmental Distribution and Taxonomic Diversity of Alkaliphiles. In *Extremophiles Handbook*. https://doi.org/10.1007/978-4-431-53898-1_4

**Supplementary material**

A     *L. manihotivorans* relative abundance

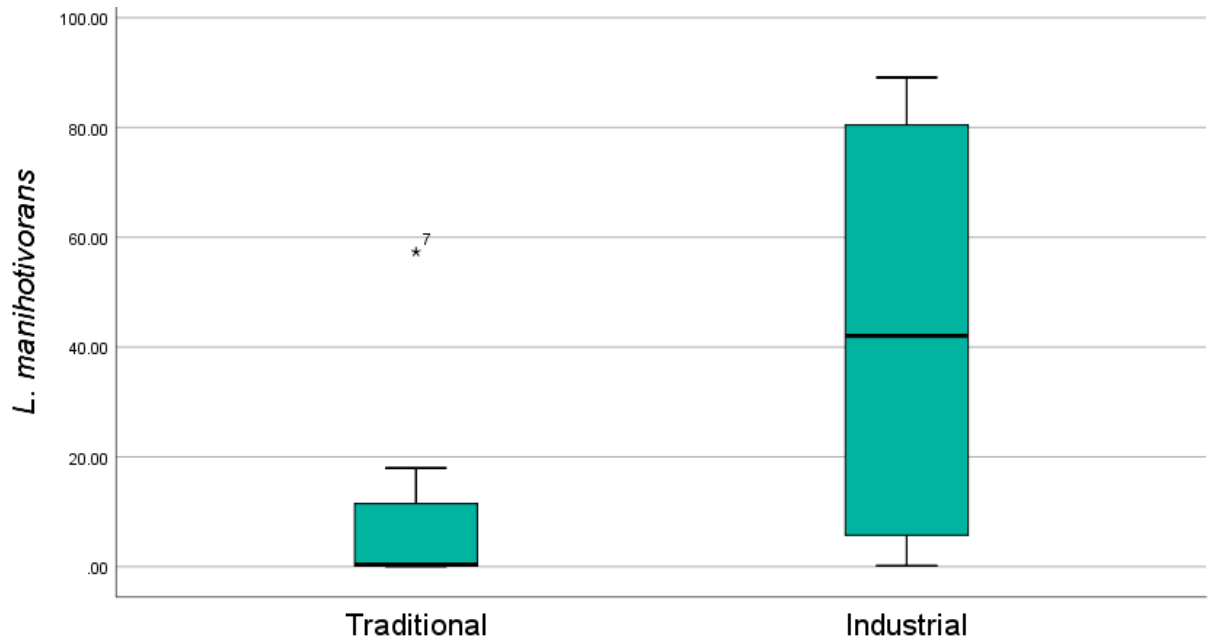

*Marinilactibacillus* relative abundance

B

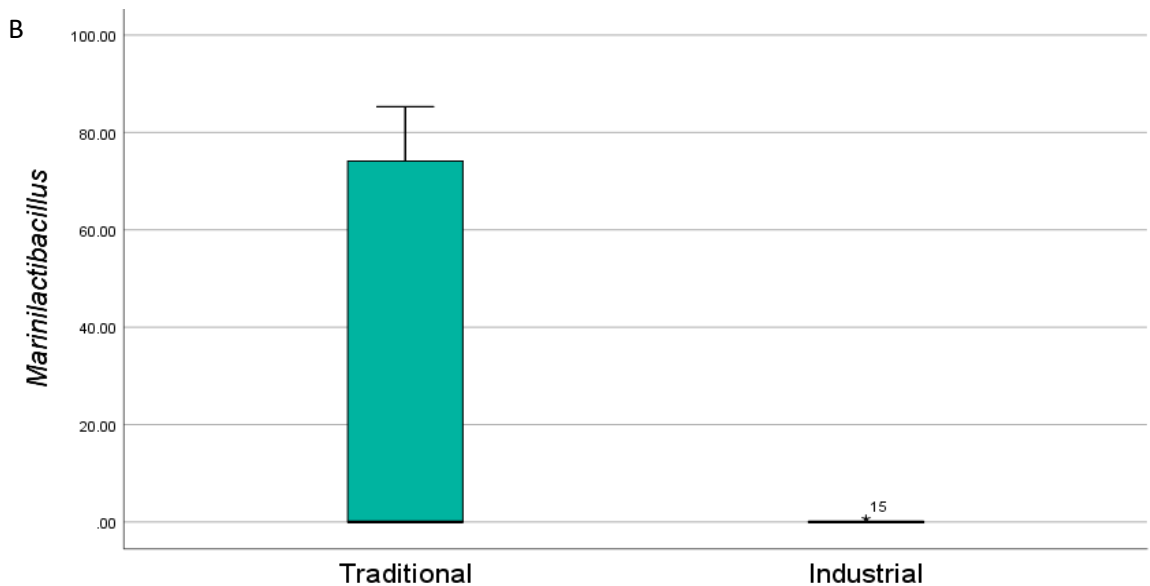

**Figure S1**. Boxplots showing the differences in the relative abundance of: a) *L. manihotivorans* (A) and b) *Marinilactobacillus* (B) between traditional and industrial Halloumi, tested with Mann-Whitney U test. (p=0.05)

**Table S1.** Sample information, microbial diversity and sequence abundance

| Sample ID | Number of reads (genus level) | Number of reads (species level) | Shannon (genus level) | Simpson (genus level) | Chao1 (genus level) | Observed OTU |
|---|---|---|---|---|---|---|
| G_1 | 168086 | 135314 | 1.74 | 3.19 | 311.07 (287.54 - 354.70) | 266 |
| G_2 | 90215 | 65241 | 2.31 | 6.05 | 340.6 (325.46 - 369.66) | 212 |
| S_1 | 106759 | 102852 | 2.21 | 5.38 | 373.76 (357.79 - 404.07) | 170 |
| GS_1 | 59520 | 47350 | 2.53 | 6.89 | 439.44 (418.47 - 477.65) | 409 |
| GS_2 | 66230 | 56802 | 2.75 | 8.59 | 494.71 (471.94 - 536.03) | 399 |
| GS_3 | 308715 | 266029 | 2.35 | 5.12 | 504.18 (484.92 - 539.61) | 149 |
| GS_4 | 95692 | 85047 | 2.36 | 5.38 | 506.3 (488.60 - 539.19) | 183 |
| GS_5 | 16485 | 12541 | 2.36 | 5.31 | 510.48 (492.34 - 544.03) | 257 |
| GS_6 | 625388 | 577516 | 1.8 | 2.69 | 564.77 (541.52 - 606.95) | 158 |
| GS_7 | 53204 | 45889 | 1.83 | 2.81 | 558.35 (538.47 - 595.05) | 185 |
| GS_8 | 55645 | 42484 | 1.83 | 2.77 | 561.23 (542.28 - 597.04) | 258 |
| GS_9 | 34768 | 32512 | 1.81 | 2.72 | 559.12 (541.06 - 593.45) | 168 |
| GSC_1 | 171937 | 160427 | 1.73 | 2.49 | 560.99 (547.97 - 587.08) | 213 |
| GSC_2 | 122019 | 107065 | 1.67 | 2.36 | 565.01 (551.51 - 592.03) | 178 |
| GSC_3 | 153712 | 138488 | 1.65 | 2.29 | 571.11 (556.76 - 599.43) | 192 |
| GSC_4 | 69501 | 44268 | 1.68 | 2.32 | 606.47 (588.15 - 641.47) | 435 |

| | | | | | |
|---|---|---|---|---|---|
| GSC_5 | 62298 | 19919 | 1.77 | 2.45 | 611.22 (592.07 - 647.77) | 248 |
| GSC_6 | 503953 | 471625 | 1.57 | 2.08 | 615.16 (595.57 - 652.54) | 142 |